\documentclass[num-refs]{wiley-article}

\usepackage{graphicx}
\usepackage[space]{grffile}
\usepackage{latexsym}
\usepackage{textcomp}
\usepackage{longtable}
\usepackage{tabulary}
\usepackage{booktabs,array,multirow}
\usepackage{amsfonts,amsmath,amssymb}
\usepackage{natbib}
\usepackage{url}
\usepackage{hyperref}
\hypersetup{colorlinks=false,pdfborder={0 0 0}}
\usepackage{etoolbox}
\makeatletter

\newcommand{\ket}[1]{\lvert #1 \rangle}
\newcommand{\qprod}[2]{ \langle #1 | #2 \rangle} 
\newcommand{\braopket}[3]{\langle #1 | #2 | #3\rangle} 

\newcommand{\subf}[2]{%
  {\small\begin{tabular}[t]{@{}c@{}}
  #1\\#2
  \end{tabular}}%
}

\makeatother
\newif\iflatexml\latexmlfalse

\AtBeginDocument{\DeclareGraphicsExtensions{.pdf,.PDF,.eps,.EPS,.png,.PNG,.tif,.TIF,.jpg,.JPG,.jpeg,.JPEG}}

\usepackage[utf8]{inputenc}
\usepackage[english]{babel}

\usepackage{siunitx}

\iflatexml

\else

\paperfield{Quantum Computing}
\corraddress{Peter Graf, PhD, Computational Science Center, National Renewable Energy Laboratory (NREL), Golden, CO , 80401, USA}
\corremail{peter.graf@nrel.gov}
\presentadd{Computational Science Center, National Renewable Energy Laboratory (NREL), Golden, CO , 80401, USA}
\fundinginfo{NREL Laboratory Directed Research and Development program}
\fi

\papertype{Original Article}

\title{Exploring the scaling limitations of the variational quantum eigensolver with the bond dissociation of hydride diatomic molecules}

\author[1]{Jacob M. Clary}
\author[2]{Eric B. Jones}
\author[1]{Derek Vigil-Fowler}
\author[3]{Christopher Chang}
\author[1]{Peter Graf}
\affil[1]{National Renewable Energy Laboratory}
\affil[2]{ColdQuanta}
\affil[3]{Amazon Web Services}

\runningauthor{Peter Graf}

\begin{document}

\maketitle
\selectlanguage{english}
\begin{abstract}
Materials simulations involving strongly correlated electrons pose fundamental challenges to state-of-the-art electronic structure methods but are hypothesized to be the ideal use case for quantum computing algorithms. 
To date, no quantum computer has simulated a molecule of a size and complexity relevant to real-world applications, despite the fact that the variational quantum eigensolver (VQE) algorithm can predict chemically accurate total energies. 
Nevertheless, because of the many applications of moderately-sized, strongly correlated systems, such as molecular catalysts, the successful use of the VQE stands as an important waypoint in the advancement toward useful chemical modeling on near-term quantum processors. 
In this paper, we take a significant step in this direction. We lay out the steps, write, and run parallel code for an (emulated) quantum computer to compute the bond dissociation curves of the TiH, LiH, NaH, and KH diatomic hydride molecules using the VQE. TiH was chosen as a relatively simple chemical system that incorporates d orbitals and strong electron correlation.
Because current VQE implementations on existing quantum hardware are limited by qubit error rates, the number of qubits available, and the allowable gate depth, recent studies using it have focused on chemical systems involving s and p block elements. 
Through VQE + UCCSD calculations of TiH, we evaluate the near-term feasibility of modeling a molecule with d-orbitals on real quantum hardware.
We demonstrate that the inclusion of d-orbitals and the use of
the UCCSD ansatz, which are both necessary to capture
the correct TiH physics, dramatically increase the cost of
this problem.
We estimate the approximate error rates necessary to model TiH on current quantum computing hardware using VQE+UCCSD and show them to likely be prohibitive until significant improvements in hardware and error correction algorithms are available. 

\end{abstract}

\section{Introduction}
\label{sec:intro}
To accelerate widespread decarbonization
 it is generally acknowledged that improved materials  and chemicals  have a large role to play and that exploration 
 and design via simulation can be extremely valuable. 
This is the case for batteries, photovoltaics, carbon capture, utilization and storage, catalysis, etc.
Because many of the systems in question require high-fidelity electronic structure calculations that can become extremely computationally expensive for classical computers, quantum chemists are increasingly interested in the prospects for quantum computing to accomplish these simulations.

A standard example and unsolved challenge is understanding the naturally occurring nitrogenase enzyme, which allows for nitrogen fixation under ambient conditions. In contrast, the current industrial equivalent to this enzyme is the highly energy intensive Haber-Bosch process, which alone accounts for between one and two percent of  global carbon emissions and energy usage \cite{ghavam2021sustainable,capdevila2019electrifying}.
Consequently, an improved understanding of the nitrogenase enzyme and other naturally occurring catalysts may allow for significant developments in 
carbon free, energy efficient industrial processes.

Unfortunately, a detailed understanding of the nitrogenase catalytic site (FeMoco) is complicated by the high degree of electron correlation present in its electronic structure. This characteristic necessitates the use of the most computationally expensive electronic structure methods and greatly limits studies of FeMoco to only the smallest system sizes. 
Note, too, that FeMoco is just an examplar; it is one of many difficult challenges in
computational catalysis specifically and quantum chemistry generally. To achieve decarbonization and other societal goals relying on advanced chemistry we may
require a fundamentally more efficient means of computing important materials and chemistry properties.

Quantum computing, in principle, by overcoming the scaling limitations of classical computing, offers just such a game-changing paradigm shift. 
But while it is now well established that real quantum computing 
hardware can be used to simulate relatively simple chemical systems, there are still no known use cases where a quantum computer has simulated something that could not be simulated on a classical computer, much less so for a chemistry with practical applications. In this work, we discuss how this divide might be bridged by focusing on quantum computing simulations of a molecule that begins to capture the complexity present in transition metal systems. 

\subsection{Classical quantum chemistry}
The ability to predict the physical properties of molecular and extended systems with chemical accuracy (resolution of 1 kcal/mol) using quantum chemical techniques has long been a driving goal for computational research. Although many methods have been developed, few entirely ab-initio approaches are able to make consistently chemically accurate predictions for metallic and semiconducting molecules, bulk 
phases, and surfaces. Coupled cluster accounting for connected single, double, and triple excitations (CCSD(T), where the triple 
excitations are accounted for perturbatively) and full configuration-interaction (FCI) are among the highest accuracy approaches.
However, both of these approaches scale steeply with system size and can only currently be used to model small molecular systems. 
For a system of $n$ electrons, CCSD(T) scales as $n^7$ while FCI scales as $n!$, meaning FCI is prohibitive beyond calculations with about 20 electrons in 20 molecular orbitals \cite{lehtola2017cluster,booth2010approaching,vogiatzis2017pushing}.

\subsection{Quantum computing and VQE}
 Quantum computers can potentially  overcome this severe scaling due to their ability to 
 simultaneously represent and manipulate a linear combination of $2^n$ states on $n$ qubits \cite{mcardle2020quantum}. More specifically, the 
 variational quantum eigensolver (VQE) framework uses a quantum computer to prepare a parameterized wavefunction and measure fixed-accuracy 
 expectation values of the many-body Hamiltonian while a classical optimizer iteratively updates the wavefunction 
parameters \cite{mcardle2020quantum,cao2019quantum}. The VQE is variational, thus in principle it allows for iterative improvement in 
 the prediction of a chemically accurate energy and other system observables while simultaneously exhibiting only linear scaling with 
 respect to the number of qubits. Despite the significant advantages of this approach, the VQE has currently only been used on 
 relatively small chemical systems to predict properties such as bond lengths or reaction barriers, primarily due to the limitations of 
 current quantum hardware \cite{o2016scalable}. Among the largest systems currently modeled on a quantum computer without using embedding techniques 
 are $H_x$ chains (where $x$ is the number of H atoms), alkali hydride diatomic molecules, and $N_2H_2$ 
 \cite{o2016scalable,kandala2017hardware,mccaskey2019quantum,google2020hartree}. Additionally, current hardware limitations and/or the 
 use of embedding techniques frequently necessitate orbital down-selection in which only orbitals near the Fermi level are represented 
 on a quantum computer, with the contribution of the rest determined classically using Hartree-Fock theory. Ideally, quantum computers 
 will become large enough for orbital-down-selection to be rendered unnecessary. Even if orbital down-selection is used, however, the 
 significantly improved scaling behavior exhibited by quantum computers will still allow for larger active spaces to be chosen. In 
 general, the VQE is highly flexible and allows one to account for the specifications of quantum hardware, thus seeing its application 
 in a wide variety of studies \cite{cao2019quantum,o2016scalable,kandala2017hardware,mccaskey2019quantum,google2020hartree,ryabinkin2018constrained,parrish2019quantum,nakanishi2019subspace,kokail2019self}.

\subsection{TiH and paper outline}
Previous quantum computing resource assessments suggest that molecules such as FeMoco are well out of reach of current quantum computing hardware \cite{reiher2017elucidating}. 
As our goal is not just resource assessment but construction and execution of actual VQE calculations, we selected an intermediate level of chemical difficulty that we suggest a priori is likely  out of reach for current quantum computing hardware and algorithms, but is also possibly attainable in the not too distant future.
 We will probe the scaling of the VQE approach on several hydride diatomic molecule systems, specifically LiH, NaH, KH, and TiH. While these are all of interest in order to understanding the scaling of the VQE, TiH is the main target. TiH was selected because it provides both an approximate model for a bond that may form during a variety of catalytic reactions and renewable energy technologies \cite{bauschlicher1988full,panayotov2017catalysis} while also being one of the simplest chemical systems containing d-electrons. The partially filled d-orbitals allow for multiple electron configurations, which is a common feature for systems difficult to model with a classical computer. The presence of multiple configurations allows us to study how this complexity propagates within the VQE as well. 

The main contributions of this paper are as follows:
\begin{itemize}
\item{We develop a software framework to compute electronic structure and bond dissociation curves using VQE for relatively large systems
that benefit from, e.g., more efficient Hamiltonian measurement using relationships between commuting Pauli strings, optimization of Pauli string term consolidation, and measurement parallelism. We demonstrate this pipeline using an emulated error-free quantum devices on a classical supercomputer and
compute the properties of several hydride diatomic molecules. These calculations provide practical insights into the difficulty in scaling the VQE algorithm into larger chemical systems.  }
\item{We  compute the fidelity that would result on a real quantum computer with a range of error rates and confirm that although simple molecules can already be simulated on real quantum hardware, the possible system sizes are still small enough to also be modeled using classical computers. For molecules of more practical interest for catalysis,
there is still a wide gap between current hardware and required error rates.}
\item{We discuss several subtleties that are encountered when moving from proof-of-concept studies to real-world applications. 
For example, we describe in detail how basis set choice can affect not only the ground state energy but the ground state configuration itself,
and that these assessments increase in difficulty when the corresponding classical calculations are not feasible.}
 \end{itemize}

Most generally, this paper seeks to begin building a bridge from the current state of VQE to realistic applications in catalysis. We show that the above additions to the VQE can together be leveraged to allow quantum chemical calculations for systems on the boundary of accessibility.

 \section{Results and Discussion}
 \label{sec:results}
 The following sections describe the VQE formulation, then what we learn about hydride electronic structure from classical calculations.
 Next we discuss efficient parallel Pauli string measurement and optimization, followed by the results of 
  error-free, emulated, quantum computations for the hydride systems.  Finally, we estimate the fidelity of these calculations on real quantum hardware in order to predict the likelihood of successfully completing these calculations in the presence of noise.
While we do not perform any computations on real quantum computing hardware, the primary focus of this study is instead to develop the code, perform relevant calculations 
on an error-free quantum computer emulated on a classical supercomputer, and then discuss, given
realistic hardware error rates,  what \emph{would} have happened had we used real quantum computing hardware. As a result, this work aims to provide a realistic assessment of running these calculations on existing devices while developing a workflow that can be deployed on real quantum computing hardware as algorithmic advances stabilize and hardware error rates decrease.
  
 \subsection{VQE formalism}
 \label{sec:formalism}
 Within the Born-Oppenheimer approximation, a chemical system is described as electrons interacting in the potential produced by the 
 atomic nuclei at fixed positions. The Hamiltonian, $H$, of each system can be written in various forms. In this work, we use the 
 so-called second-quantized Hamiltonian form, where systems are described using empty or occupied single-particle spin orbitals and 
 interactions between electrons are represented using creation and annihilation operators. This form of the electronic Hamiltonian is 
 written as Eq. (\ref{eq:ham}): 
\begin{equation}   H = \sum_{p,q}{h_{pq}a^{\dagger}_p  a_q}  + \frac{1}{2} \sum_{p,q,r,s} { h_{pqrs}a^{\dagger}_p a^{\dagger}_q  a_r a_s,         }    
\label{eq:ham}
\end{equation}
 where
$$ h_{pq}  =   \int {dx  \phi^*_p(x) \left( - \frac{\nabla^2}{2}  - \sum_I \frac{Z_I}{|r-R_I}  \right) \phi_q(x) }, $$
and
$$ h_{pqrs}   = \int {dx_1 dx_2} \frac{\phi^*_p(x_1) \phi^*_q(x_2)  \phi_r(x_2)  \phi_s(x_1)}  { |r_1 - r_2 | }.         $$
Because the Coulomb interaction between electrons is a two-body interaction, the terms of this Hamiltonian contain up to two creation 
 and two annihilation operators. The integral for $h_{pq}$ describes the kinetic energy terms of electrons and their Coulomb interaction 
 with the nuclei while the integral for $h_{pqrs}$ describes the electron-electron Coulomb repulsion. 
Next, the second quantized Hamiltonian with operators acting on indistinguishable Fermions is mapped to operators acting on 
distinguishable qubits. The result of this mapping is a linear combination of products of single-qubit Pauli operators, with each 
product called a Pauli string. Although various encoding schemes exist, we used the parity encoding scheme because it mapped the 
Hamiltonians of the hydrides using the fewest number of Pauli strings as compared to the Jordan-Wigner and Bravyi-Kitaev schemes \cite{seeley2012bravyi}. To 
calculate expectation values of the mapped Hamiltonians, we used the chemically inspired unitary coupled cluster ansatz which is 
truncated to either single excitations (UCCS) or singles and doubles excitations (UCCSD), to represent our trial wavefunction. Note that
UCCS is a cheaper but fundamentally less accurate method than UCCSD, as the doubles amplitudes capture the electron correlation and the singles amplitudes 
mostly account for relaxation effects.
The ground state can be calculated from the Hartree-Fock reference state, $\ket{\Psi_{HF} }$, using excitation operators. The UCCSD ansatz can 
thus be written as:
\begin{equation}
\ket{\Psi_{UCCSD}(\Theta)} = \exp^{ T(\Theta) - T^{\dagger}(\Theta)} \ket{\Psi_{HF}}
\end{equation}
where $T(\Theta)=T_1 (\Theta_1 )+T_2 (\Theta_2)$ is the cluster operator, which is expanded using the connected operators $T_1 (\Theta_1 
)$ and $T_2 (\Theta_2)$ in order to introduce singles and doubles excitations into the wavefunction, respectively, and $\Theta$ is a 
vector of parameters needed to specify the single- or two-qubit unitary gates in the quantum circuit \cite{cao2019quantum}. 

Finally, the ground state of each Hamiltonian is found by determining the set of parameters that minimize energy expectation value, as 
described by the variational principle  
\begin{equation}
    E_0 \le \frac{\braopket{\Psi(\Theta)}{H}{\Psi(\Theta)}}{\qprod{\Psi(\Theta)}{\Psi(\Theta)} }, 	
\label{eq:variational}
\end{equation}
where $E_0$ is the true ground state energy of $H$. The state preparation and measurement of the quantum circuit is performed on the 
quantum computer. The transformation of the quantum circuit statevector into an expectation value and the subsequent parameter optimization are 
both performed on a classical computer. 


\subsection{Hydride electronic structure}
\label{sec:hydride}
LiH, NaH, and KH are all diatomic molecules composed of an alkali earth metal atom and a hydrogen atom. These molecules were selected because of their relatively few 
valence electrons and lack of d-orbitals participating in their bonding. All three molecules adopt similar orbital occupations with the highest energy s orbital on Li, 
Na, or K bonding with the H 1s orbital to form a fully filled valence shell. As these diatomic molecules dissociate, the two asymptotic limits of the diatomic molecules 
(bound and dissociated) will mix. Despite this, the VQE is still able to predict chemically accurate total energies for these molecules \cite{o2016scalable,mccaskey2019quantum}


In contrast to the bound alkali hydride diatomic molecules, bonding in TiH and other transition metal hydride diatomic molecules can arise from 
multiple occupations that have been studied both experimentally and computationally \cite{bauschlicher1988full,walch1983casscf,chong1986theoretical}. The two primary occupations to consider are 
the $3d^3 4s^1$ and $3d^2 4s^2$ occupations \cite{bauschlicher1988full}. In the first occupation, a 4s-1s bond is formed between Ti and H, leading to the $^4\Phi (S = 4)$ state with an occupation 
of $...6\sigma^2 7\sigma^1 3\pi^1 1\delta^1$, where the $6\sigma$ orbital is the Ti-H bonding orbital, the $3\pi$ and $1\delta$ orbitals are the Ti 3d-like orbitals, and the $7\sigma$ orbital is 
a mixture of the 4s, 4p, and 3d orbitals. In the second occupation, hybridization of the 4s-4p orbitals and 4s-3d orbitals can occur, with one hybrid orbital bonding with H. The $^2\Delta (S = 2)$
state has an occupation of $...6\sigma^2 7\sigma^2 1\delta^1$, where the $6\sigma$ orbital is the Ti-H bonding orbital and the $7\sigma$ orbitals is the nonbonding 4s-3d hybrid orbital. Because 
Ti has few d-electrons, the Ti 4s and 3d orbitals are spatially similar, and Ti-H bonding via the 4s-3d hybrid orbital competes with Ti-H bonding via the 4s-4p hybrid orbital such that the higher 
energy $^2\Delta (S = 2)$ state lies only 0.011 Ha above the $^4\Phi (S = 4)$ state according to FCI calculations \cite{bauschlicher1988full}. We note that the same asymptotic mixing that occurs 
during alkali hydride diatomic molecule dissociation is also present during TiH dissociation. 

\begin{figure}
        \centering
            \includegraphics[width=\linewidth]{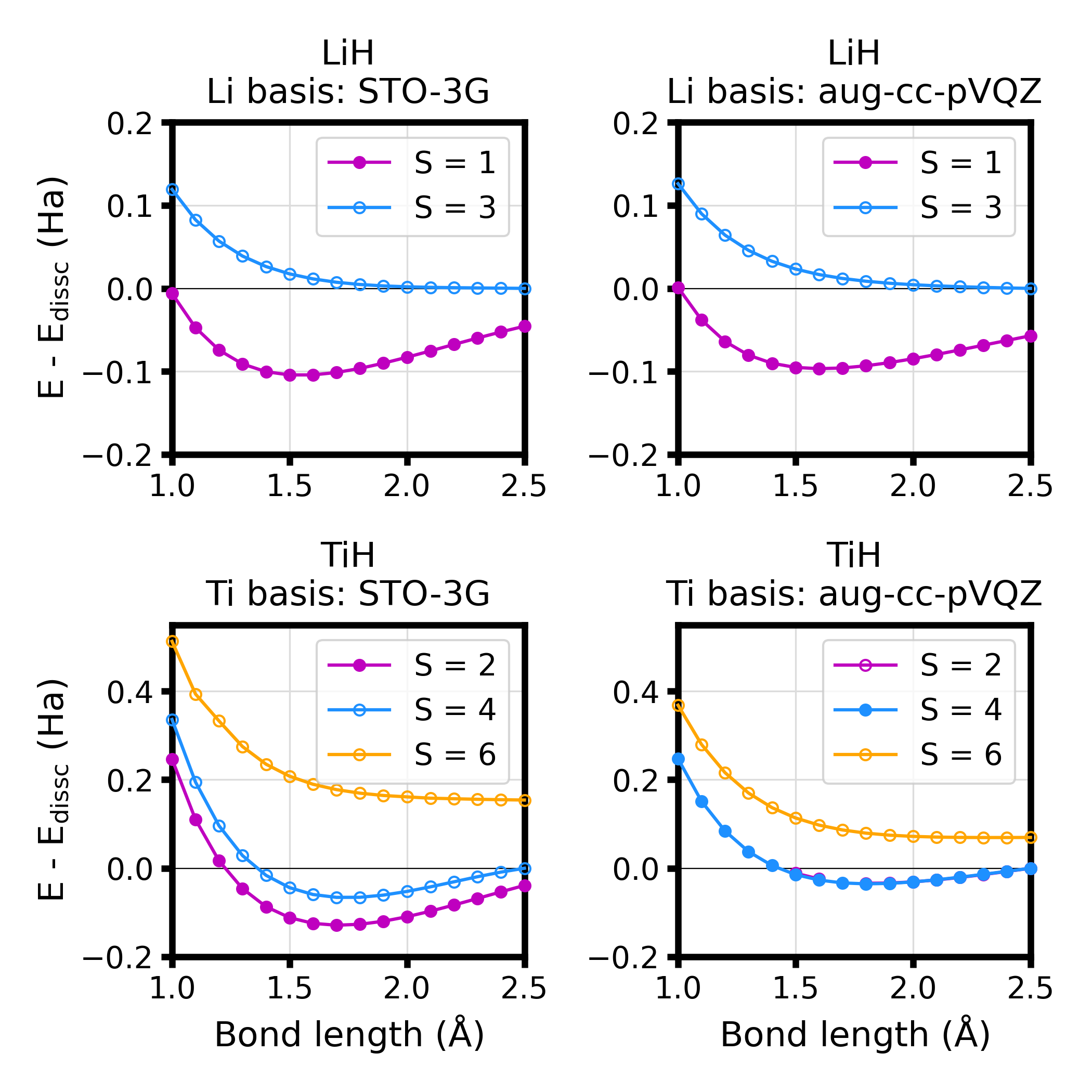}
        \caption{Predicted CCSD bond dissociation curves of the LiH and TiH diatomic molecules initialized to different spin multiplicities using different orbital basis sets on a classical computer. The predicted TiH ground state configuration changes depending on the orbital basis set chosen. The ground state configuration is denoted by filled markers while higher energy configurations are denoted by empty markers.}
        \label{fig:hydride}
\end{figure}

The choice of basis set used to model TiH can change the predicted ground state configuration. Classical CCSD quantum calculations were used to calculate the total energies of LiH and TiH initialized to different spin multiplicities (see Figure \ref{fig:hydride}). These calculations show 
that using the STO-3G basis set on the Ti atom results in the low-spin configuration being the ground state configuration by 0.063 Ha while
using the aug-cc-pVQZ basis set on the Ti atom predicts the correct opposite configuration ordering, with the high spin configuration 
being 0.001 Ha more stable. This issue is of course not present in LiH because the only other possible spin multiplicity does not have an 
energy minimum over the same range of bond lengths. As one would expect, these results demonstrate that the choice of orbital basis set can 
be an important consideration for systems with multiple possible spin configurations and may qualitatively affect the predicted system properties. 
Unfortunately, the capability of current quantum computers greatly limits the choice of orbital basis set because this choice can drastically 
impact the computational cost of the computations, as discussed further in section \ref{sec:parmeas}. As a result, while basis set choice remains a key 
cost parameter in VQE calculations, care must be taken that a specific choice does not alter the fundamental properties of the system of 
interest. In the short term, results from existing classical calculations, including HF, post-HF, and DFT methods can provide guidance on 
the impact of basis set choice for classically studied chemical systems. In the long term, VQE calculations without corresponding classical 
results should be converged with respect to basis set choice, as is currently commonly done for classical calculations.

\begin{figure}
\centering
\begin{tabular}{cc}
\subf{\includegraphics[width=60mm]{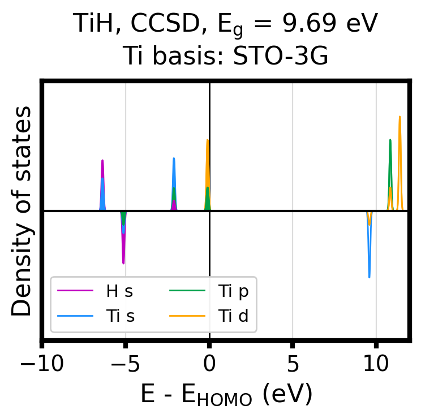}}
     {TiH CCSD PDOS calculated with the STO-3G basis set on Ti}
&
\subf{\includegraphics[width=60mm]{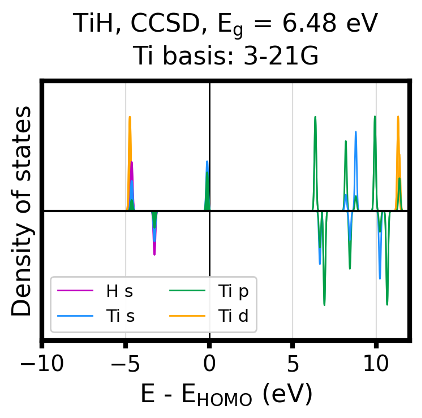}}
     {TiH CCSD PDOS calculated with the 3-21G basis set on Ti}
\\
\subf{\includegraphics[width=60mm]{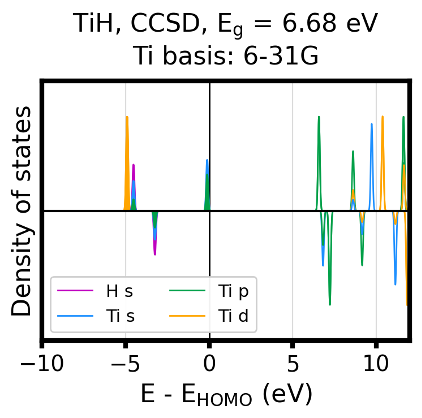}}
     {TiH CCSD PDOS calculated with the 6-31G basis set on Ti}
&
\subf{\includegraphics[width=60mm]{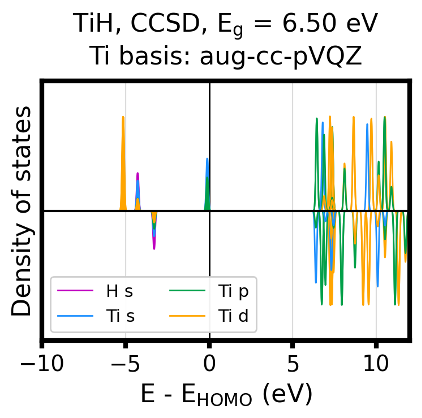}}
     {TiH CCSD PDOS calculated with the aug-cc-pVQZ basis set on Ti.}
\\
\end{tabular}
\caption{Selected PDOS for the studied hydride diatomic molecules using CCSD on a classical computer.
The STO-3G basis set was always used for H. The bonding/orbital hybridization mentioned previously is 
visible in the valence region of the PDOSs.}
\label{fig:pdos}
\end{figure}

In addition to the correct energy ordering of different molecular spin configurations, basis set choice must also allow for an 
accurate description of a chemical system’s electronic structure, with the frontier orbitals being of particular importance. For 
LiH, classical CCSD calculations at the experimental Li-H bond length using the STO-3G basis set predict frontier orbitals with 
similar degrees of Li s, Li p, and H s character as calculations using the aug-cc-pVQZ basis set. However, the STO-3G basis set 
still predicts a LiH HOMO-LUMO gap, $E_g$, of 9.90 eV, while the aug-cc-pVQZ basis set predicts a gap of only 7.99 eV (Figure \ref{fig:ccsdAH}). 
To further test the $E_g$ and frontier orbital character dependence for these molecules on basis set choice, we repeated the above 
calculations using a larger variety of basis set choices (STO-3G, 3-21G, 6-31G, cc-pVDZ, aug-cc-pVDZ, cc-pVQZ, aug-cc-pVQZ). These 
calculations show that the 3-21G basis set generally allows for $E_g$ and frontier orbital character predictions that approximate 
the results calculated using the aug-cc-pVQZ basis set for LiH, NaH, and KH (Figures \ref{fig:ccsdHomoLumo} and \ref{fig:ccsdProj}). 
For $^4\Phi (S = 4)$
TiH, $E_g$ calculated 
using the 3-21G basis set was much closer to the aug-cc-pVQZ $E_g$ than the STO-3G $E_g$ was (Figure \ref{fig:ccsdHomoLumo}). However, Figure \ref{fig:pdos} shows that 
the STO-3G, 3-21G, and 6-31G basis sets all predicted that the minority spin LUMO orbitals were higher in energy than any of the 
correlation consistent basis sets, resulting in changes to the predicted LUMO characters.

As is well-known in classical calculations, the cost of total energy minimization using the VQE scales 
with the number of orbitals available for occupation and thus electron excitation. As a result, 
significant computational resource savings may be achieved by freezing core orbitals that negligibly 
contribute to the bonding and/or removing high energy/non-bonding virtual orbitals from the active space. Indeed, 
the frozen core approach is an approximation already commonly applied with great success in both 
molecular and periodic DFT calculations \cite{patkowski2007frozen,blochl1994projector,kresse1999ultrasoft}. Classical CCSD calculations for LiH, NaH, and KH show that 
the highest energy occupied molecular orbital lies approximately 20 eV higher in energy than the next 
highest energy occupied orbital, justifying the selection of this orbital as the only valence orbital. 
$^4\Phi (S = 4)$ TiH has five valence spin orbitals located within 10 eV of the Fermi level, with relative energies 
that all depend on basis set choice. These five occupied spin orbitals can be treated as unfrozen 
valence orbitals because the next highest energy orbital is approximately 40 eV lower in energy. 

To summarize, 1) just as in classical quantum chemistry, basis set choice is critical, and validation by
convergence with respect to basis set is desirable, and 2) the TiH system exhibits exactly the type of
nuance that necessitates this type of validation.

\subsection{Efficient and parallel measurement of Hamiltonian Pauli strings}
\label{sec:parmeas}
\begin{figure}
        \centering
            \includegraphics[width=0.5\linewidth]{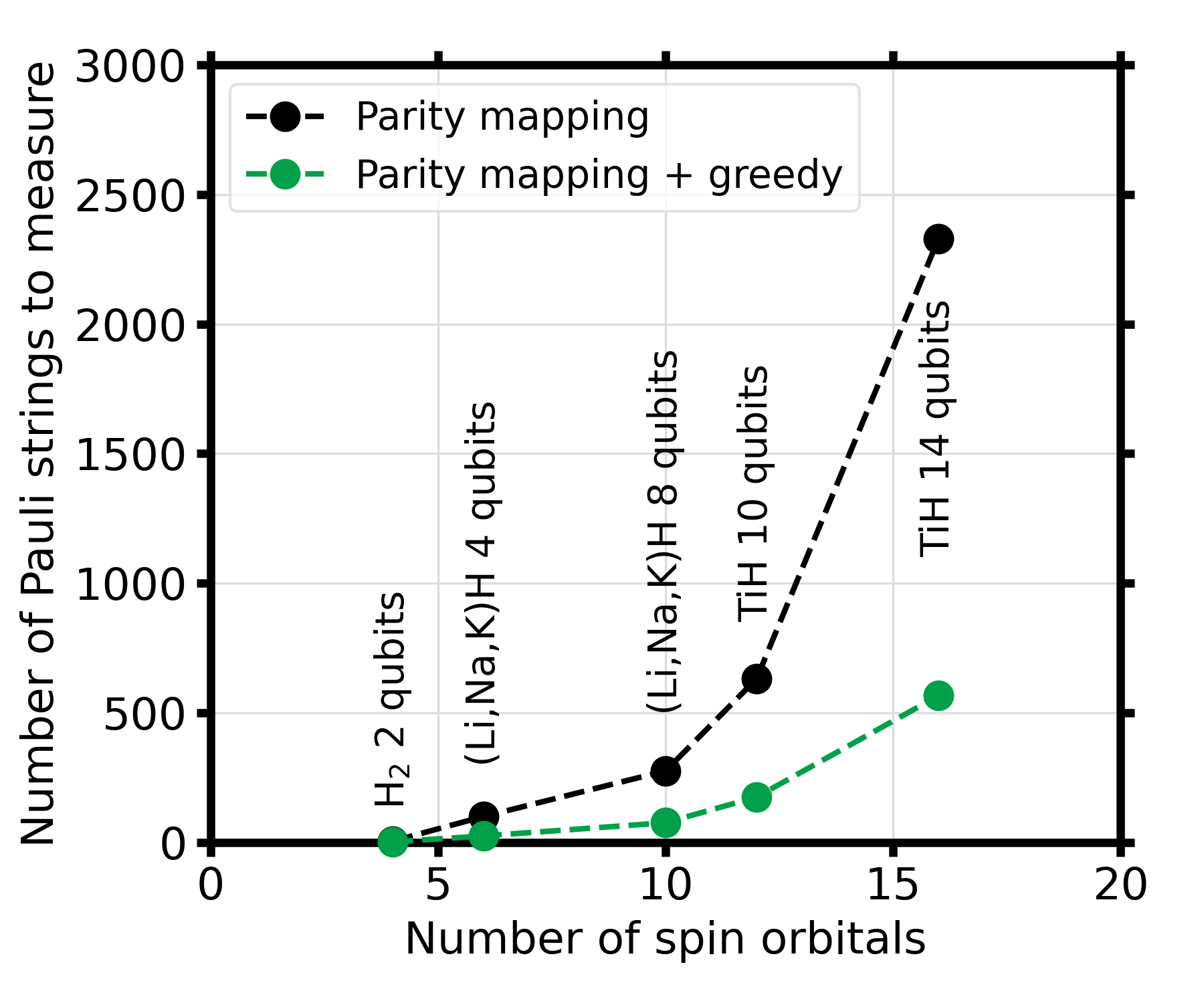}
        \caption{Scaling of number of Pauli strings in Hamiltonian for different molecules using the STO-3G basis set. The dashed line is included as a guide to the eye.}
        \label{fig:pstringscaling}
\end{figure}

The number of Pauli strings in the second-quantized chemical Hamiltonian on $n$ qubits grows as $n^4$ with 
the number of spin orbitals in the calculation (Figure \ref{fig:pstringscaling}) \cite{crawford2021efficient}. 
However, separate explicit measurements of 
the compiled quantum circuit in each set of measurement bases in the Hamiltonian is not necessary. For 
example, it is well known that the expectation value of the $H_2$ Hamiltonian within the Bravyi-Kitaev 
mapping can be measured much more efficiently than one would naively expect from the five term 
Hamiltonian \cite{o2016scalable}. 
This is because Pauli strings that share an eigenbasis can be measured simultaneously, and 
Pauli strings share an eigenbasis if and only if they commute \cite{gokhale2019minimizing}.
As a result, the Hamiltonian can be partitioned into groups of qubit-wise commuting Pauli strings, each of which 
can be measured simultaneously on a quantum computer and later used to reconstruct the expectation value of the original 
Hamiltonian on a classical computer. Without accounting for any additional symmetries in the Hamiltonian, the general 
problem of determining the smallest set of unique measurements that can be used to determine the energy of any term in the 
Hamiltonian is equivalent to the clique cover and set cover combinatorics problems, both of which are known to be NP-hard
\cite{gokhale2019minimizing,verteletskyi2020measurement,jena2019pauli}.
Briefly, the set cover 
problem can be stated as asking for the smallest number, $N_p$, of given subsets whose union equals a 
universal set, \{U\}. Thus, the exact solution for the most efficient set of measurements to make for a 
system can quickly become intractable and, ironically, could itself likely benefit from quantum 
optimization heuristics. 
One approach to approximate a solution to the set cover problem involves the use of an iterative greedy algorithm. 
This algorithm iteratively adds the measurement basis (a Pauli string defining measurement bases for all qubits) 
to the final set of measurement bases \{B\},  that commutes with the greatest remaining number of Pauli strings in \{U\} (the Hamiltonian)
i.e., that covers the greatest number of Pauli strings that have yet to be covered. The commuting family of Pauli strings 
are then removed from \{U\} prior to the next iteration.
We found that  this algorithm reduces the scaling prefactor for this problem by approximately a factor of five for the 
systems studied here and thus significantly decreases the number of measurements required for a given Hamiltonian (Figure \ref{fig:pstringscaling}). Nevertheless, as the sophistication of the chosen basis set increases, the number of Pauli strings needed to describe the Hamiltonian and the number of possible orbitals that can be included in the active space grows. For example, TiH modeled using 14 qubits and only the 6-31G basis requires about 2.5x the original number of Pauli strings as the STO-3G basis set. 

Although efficient measurement of Hamiltonian Pauli strings can decrease the number of measurements 
required for large Hamiltonians, a full optimization of the TiH wavefunction using this approach can 
still involve optimization of dozens to hundreds of parameters, $\Theta$, primarily depending on the 
number of orbitals included in the calculation. However, further savings can be had because calculating 
the expectation value of the quantum circuit in each set of measurement bases can be performed 
independently.  As a result, parallelization over $P$ processors of the energy evaluation measurements of 
bases in \{B\} allows VQE minimization to become accessible for larger chemical systems. We emphasize that 
this approach is primarily relevant for the study of problems too large to reliably model on existing 
quantum hardware at existing error rates and thus necessitates the use of emulation on classical 
computers in a noiseless simulation environment. Secondarily, this approach is useful for the study of 
either general or system-specific VQE optimizations prior to calculations on real quantum devices.

\begin{figure}
        \centering
            \includegraphics[width=0.5\linewidth]{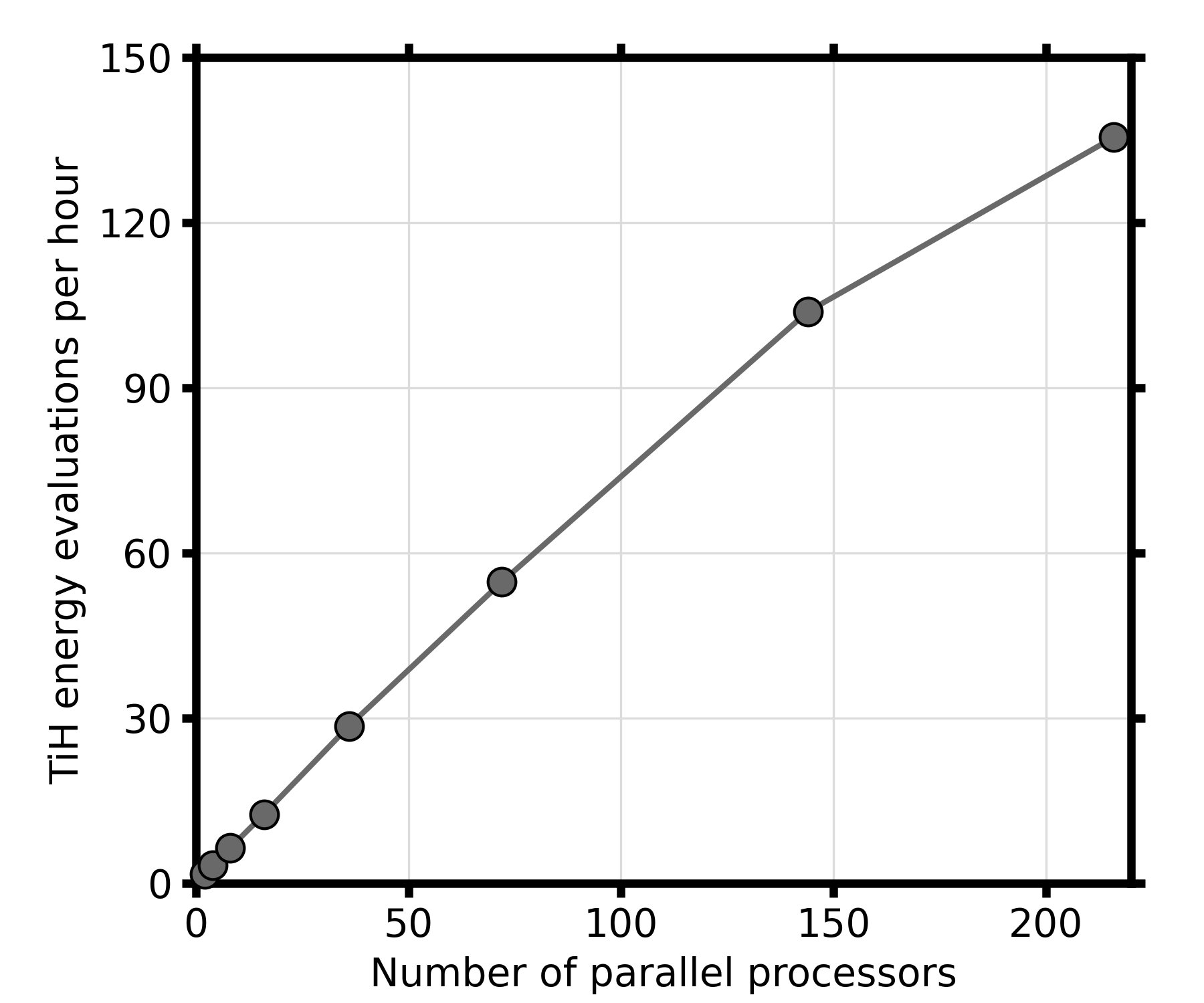}
        \caption{TiH energy evaluations per hour as a function of the number of processors the basis set measurements are parallelized over. Approximately linear scaling is observed until the number of processors exceeds the number of Pauli strings in the Hamiltonian. }
        \label{fig:evalsperhour}
\end{figure}

In this scheme, the set of final measurement bases is distributed across all processors such that each 
processor collects statistics for $\lceil N_p/P \rceil $ measurement bases on a local quantum circuit (see Methods). 
These statistics are then aggregated across all processors such that the expectation value of any Pauli 
string in the original Hamiltonian can be reconstructed from these statistics. On dual-socket nodes with 
two 3.0 GHz, 18-core Intel Xeon Gold 6154 Skylake processors, we find that the time required per energy 
evaluation can be decreased by at least an order of magnitude and can allow the optimization problem to become 
computationally feasible (Figure \ref{fig:evalsperhour}). These results demonstrate that qubit parallelization can help 
measurement throughput as is currently widely used in classical computers. Importantly however, the 
efficient selection of \{B\} and measurement parallelization only provide polynomial speed increases. The 
exponential scaling of the problem still dominates, indicating that Pauli string measurement basis 
parallelization and optimization must be married with high-fidelity quantum hardware in order to furnish 
a powerful computational paradigm for quantum chemistry generally. 

\subsection{Choice of classical optimizer}
As mentioned above, the size of the Hamiltonian, and thus the parameter space to optimize, grows 
exponentially with system size. As a result, the difficulty of the classical optimization increases 
significantly for larger chemical systems, both because of the higher dimensional parameter space and 
the difficulty in quantifying the effects of noise. Gradient descent algorithms, in particular, are 
highly affected by noisy measurements of each Pauli string because noise can both increase the 
difficulty in converging to the global minimum rather than any local minima and cause estimates of the 
gradient to vary wildly, potentially making gradient descent impossible at all. Although noise will be a 
significant hurdle in modeling systems such as TiH on near term quantum devices, here we use the 
noiseless statevector simulator to decouple the effects of noise from the difficulty of the 
optimization. 

\begin{figure}
        \centering
            \includegraphics[width=0.5\linewidth]{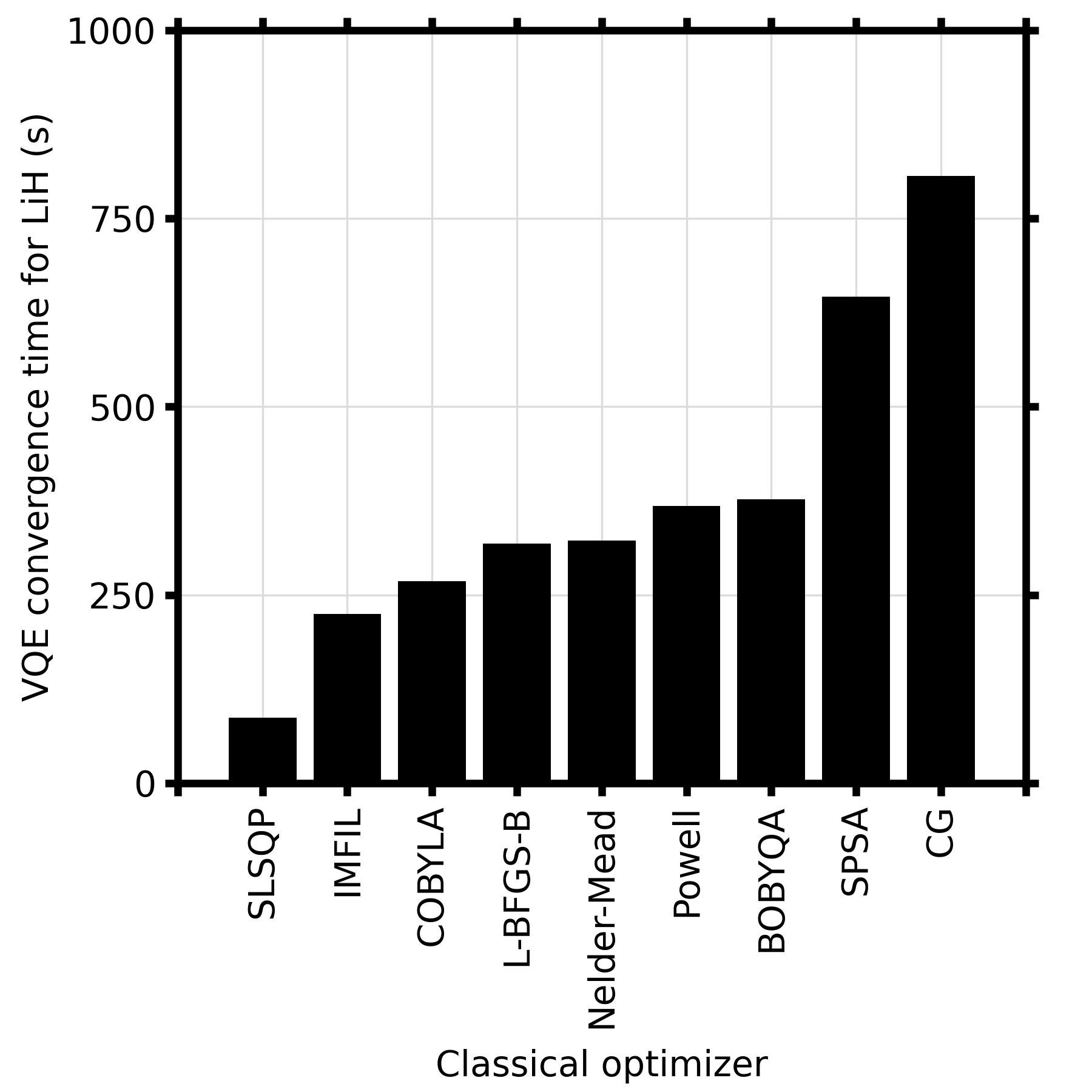}
        \caption{Comparison of optimizer performance to converge the total energy of the LiH diatomic molecule with a 6-31g basis set to within 10 meV of the exact energy.}
        \label{fig:optimizer}
\end{figure}

We study the effect of optimizer choice on the VQE time to convergence for a LiH diatomic molecule using 
the 6-31G basis set (Figure \ref{fig:optimizer}). The ideal optimizer will converge to chemical accuracy in the 
shortest time. We find that the sequential least squares programming (SLSQP) optimizer consistently 
produces the most accurate energy minimization in the fastest time \cite{kraft1988software}. However, we note that this 
optimizer is a local search algorithm that is not guaranteed to find the global minimum. This drawback 
is particularly important for systems such as TiH because the large parameter space requires an 
optimizer that can perform a broad-breadth search of the potential energy surface while still being able 
to descend into potentially narrow energy wells once they are found. Unfortunately, the classical 
optimization required for VQE is global optimization of large, nonconvex, expensive, noisy functions and 
is one of the hardest types of optimization problems there is, meaning that such an ideal optimizer does 
not exist.

\subsection{Alkali hydride diatomic molecule bond dissociation }
\begin{figure}
        \centering
            \includegraphics[width=\linewidth]{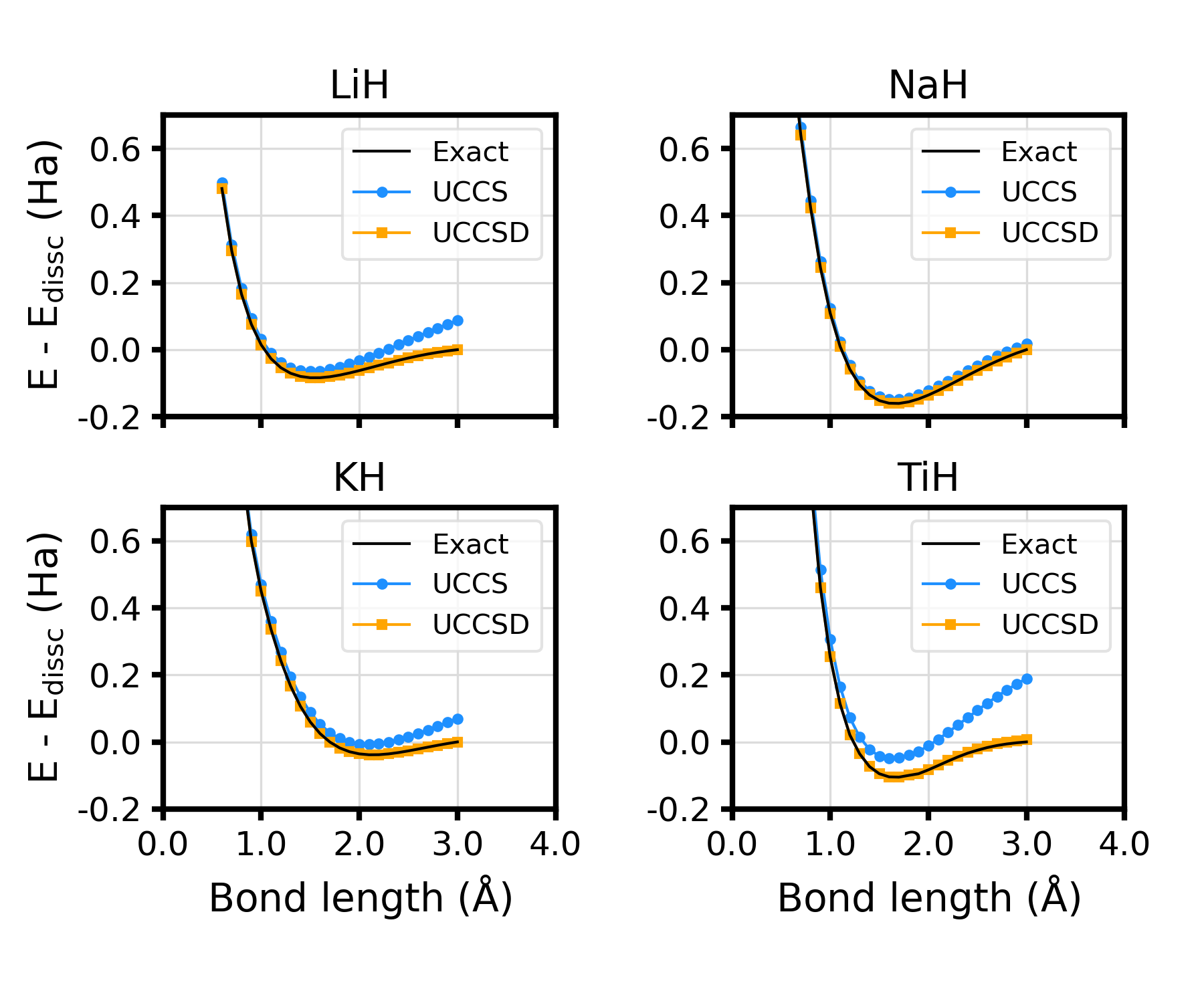}
        \caption{ Bond dissociation curves of LiH, NaH, KH, and TiH using the STO-3G basis set using the VQE+UCCS ansatz, the VQE+UCCSD ansatz, and exact Hamiltonian diagonalization.  }
        \label{fig:quantumbondissoc}
\end{figure}

We find that the VQE can predict chemically accurate bond dissociation curves for LiH, NaH, and KH using 
the STO-3G basis set (Figure \ref{fig:quantumbondissoc}) as compared to the exact bond dissociation curve (direct diagonalization 
of the Hamiltonian). Because the valence electronic structures of LiH, NaH, and KH are all extremely 
similar, the scaling of the VQE cost for these systems is similar. Furthermore, despite the much more 
severe scaling with basis set choice, the STO-3G basis set coupled with the additional orbital freezing 
and reduction discussed in section \ref{sec:hydride} also allows the TiH system to become accessible to quantum 
computer emulation schemes on reasonable timescales. We again emphasize that while important to report 
here, the physical relevance of this TiH calculation is primarily qualitative due to the inherent 
limitations in accuracy of the basis set choice, as described in Section \ref{sec:hydride}. Nevertheless, the 
performance of the VQE+UCCS(D) algorithms can still be compared to the exact dissociation curve because 
the Hamiltonian is limited by the same restrictions in all cases. We find that the UCCSD ansatz 
accurately reproduces the exact bond dissociation curve for all of the diatomic molecules. In contrast, 
the UCCS ansatz begins to significantly deviate from both the UCCSD and exact energy curves at large 
bond distances, resulting in the energy of the dissociated limit being much higher in energy. We note 
that the bond dissociation curve for TiH is very sensitive to the choice of orbitals included as the active space. 
A set that insufficiently describes the bonding may exhibit an artificial kink at the Coulson-Fischer point \cite{coulson1949xxxiv,kedziora2016bond} 
(approximately 2.4 \AA) where the bond dissociation curves for different spin configurations cross. 

\subsection{Fidelity}
Despite the usefulness of modeling chemical systems with the VQE using a noiseless emulator, ideally the 
continued improvement of real quantum device capabilities will eventually allow for the modeling of 
larger systems such as TiH. Although this work demonstrates that diatomic molecule calculations using 
the STO-3G basis set are currently accessible, TiH calculations using larger active spaces or more sophisticated basis sets can become dramatically more expensive. 

\begin{figure}
\centering
        \includegraphics[scale=0.7]{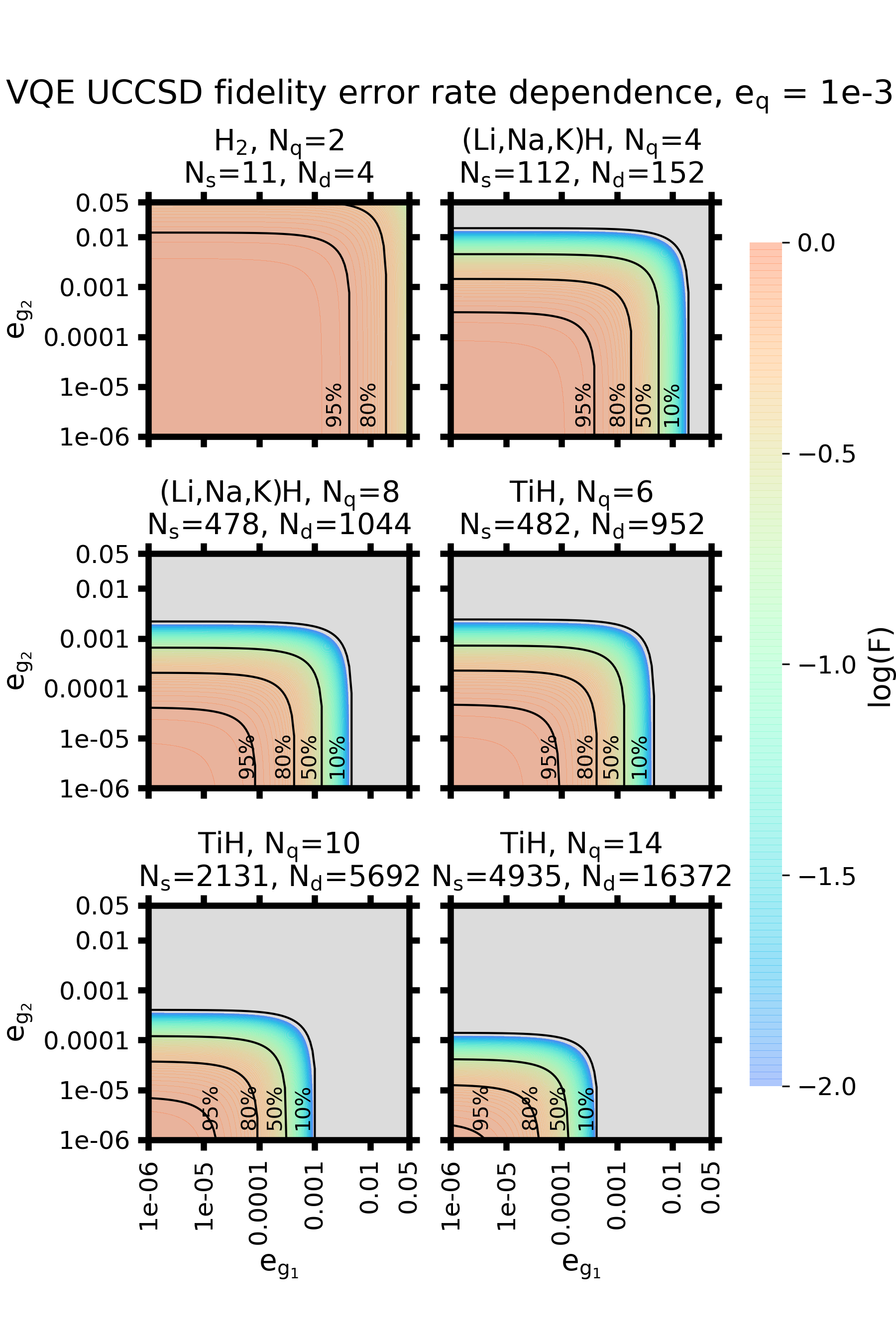}
\caption{Fidelity estimates for the UCCSD ansatz for the hydride diatomic molecules studied in this work with different numbers of qubits, single qubit gates, and two-qubit gates in the quantum circuit. A fidelity of 1.0 corresponds to no fidelity loss due to the considered errors.}
\label{fig:fidelity}
\end{figure}

In order to further understand the viability of modeling the above chemical systems on a real quantum 
device, it is useful to estimate the state preparation and measurement (SPAM), single qubit, and two-qubit 
error rates required to obtain various levels of calculation fidelity. Calculation fidelity, $F$, in 
the digital error model can be employed as a useful proxy to estimate the hypothetical VQE total energy 
calculation fidelity \cite{arute2019quantum}. To this end, $F$ was calculated using Eq. (\ref{eq:fidelity}): 
\begin{equation}
     F = (1-e_{g_1})^{G_1}  (1-e_{g_2})^{G_2} (1-e_q)^Q  ,
\label{eq:fidelity}
\end{equation}
where $e_{g_1}$ is the single qubit gate error rate, $e_{g_2}$ is the two-qubit gate error rate, 
$e_q$ is the SPAM error rate, $G_1$ is the 
number of single qubit gates, $G_2$ is the number of two-
qubit gates, and $Q$ is the number of qubits in the circuit. The error rate for each single and two-
qubit gate was assumed to be constant. Eq. (\ref{eq:fidelity}) is plotted in Figure \ref{fig:fidelity} for the circuits of the 
different hydride diatomic molecule systems discussed above using the STO-3G basis set and the UCCSD ansatz. The same results for the UCCS ansatz are shown in Figure \ref{fig:fidelitysupp}.
Existing quantum computing devices have SPAM, single, and two-qubit error 
rates of approximately 1e-2, 1e-3, and 1e-2, respectively \cite{wright2019benchmarking,rigettierror,finsterhoelzl2022benchmarking}.
As a result, Figure \ref{fig:fidelity} (bottom) shows 
that this error model predicts that the larger UCCSD scale calculations are not yet reliably 
feasible on existing hardware. We note that the fidelity of an H$_2$ molecule is predicted to be 
approximately 0.95 at current error rates (Figure \ref{fig:fidelity} top left). Both observations are consistent with the 
existing literature on chemical properties predicted with the VQE on real quantum 
computers \cite{o2016scalable,kandala2017hardware,mccaskey2019quantum,google2020hartree}.
They support the idea that our estimated gate counts and consequent fidelities provide reasonably 
accurate lower fidelity bounds for the more computationally demanding systems given the lack of 
hardware-specific circuit optimization. Without further advances in error correction schemes and 
circuit optimization for specific hardware, robust TiH models will require error rates 
approximately two orders of magnitude lower than exhibited by any existing quantum computer. In 
contrast, the UCCS ansatz might provide a much faster route towards experimental validation of 
the above results. In the near future, all of the studied hydride diatomic molecule calculations 
can likely be carried out on real quantum computers with reasonable fidelities with only an 
approximately factor of five improvement in existing error rates, however, they will still likely 
require the use of a qualitatively inaccurate wavefunction ansatz.

\section{Conclusions}
In this work, we have examined the computational complexity of VQE against a series of hydride diatomic molecules from LiH to the d-orbital-containing TiH. We show that the complexity of the 
simulation drastically increases upon incorporation of d orbitals due to more Pauli strings in the Hamiltonian, a larger number of 1- and 2-electron excitation operators, and the presence of multiple orbital occupation configurations. Furthermore, we show that although a simpler basis set facilitates VQE estimation of the TiH bond dissociation curve on current (classically emulated) quantum devices, it comes at the cost of incorrect prediction of even the ground state occupation configuration and relies on the use of error free emulation. This tradeoff will likely remain relevant within the NISQ computing era, particularly for more complicated chemical systems involving multiple transition metal atoms. 

Although the inclusion of d orbitals in the TiH VQE calculations clearly increases the model complexity, the exact impact of transition metals in a chemical system may not be as clear for more complicated chemical systems. These types of systems may lack data from prior high-fidelity classical calculations that inform future quantum computing calculations, and may instead rely on iterative VQE testing. This process, if needed, will be greatly facilitated by further developments in the VQE algorithm itself. 
 
For example, combining the UCCSD formalism with an adaptive ansatz such as ADAPT-VQE \cite{grimsley2019adaptive} has the potential to reduce the complexity of the quantum circuit (number of optimization  parameters, gate depth, etc.) significantly.  Additionally, the use of so-called transcorrelated Hamiltonians has been shown to achieve quantitatively accurate ground and excited state energies using minimal basis sets \cite{kumar2022quantum}. Optimizers that are resilient to noise will also be particularly helpful for large system modeling on real hardware \cite{muller2022accelerating}.  
Quantum embedding theory \cite{sun2016quantum} shows great potential to extend the reach of quantum computing to larger systems, particularly those with a small set of atoms that require high-fidelity methods surrounded by a less computationally complex environment. Finally, improvements in qubit error rates, error mitigation \cite{endo2018practical,kapit2017upside},
and postselection \cite{huggins2021efficient} could further improve accuracy for a given circuit depth.

It is currently unclear whether the combined improvements offered by these  developments will allow for the full modeling of molecules 
like nitrogenase on a quantum computer. We nevertheless hope that this detailed study will provide a benchmark that can be revisited following further
algorithmic and hardware developments on the way to modeling chemical systems that are both practically important and 
currently out of reach for classical computing. 

\section{Methods}
All classical CCSD electronic structure calculations were performed using the Gaussian 16 software package 
\cite{frisch2016gaussian}. The LiH, NaH, and KH diatomic molecules were modeled with 
either a spin multiplicity of 1 or 3 while TiH was modeled with spin multiplicities of 2, 4, and 6 
(section \ref{sec:hydride}) \cite{bauschlicher1988full}. The screened bond lengths were chosen to 
cover the bonding energy well present within the dissociation curve. The basis set screening was carried 
out by varying the basis set on Ti while using the STO-3G basis set on H to mimic the likely progression 
of future calculations on a quantum computer. The number of primitive gaussians used for each molecule 
calculation are shown in Figure \ref{fig:gaussians}. Altering the H basis set did not significantly 
change the qualitative energy ordering or electronic structure trends. 

Quantum computer emulation was performed using IBM's Qiskit API and simulator \cite{Qiskit}. Hamiltonian preparation 
and diagonalization and preparation of the wavefunction ansatz for each system modeled in this work were 
performed with the Qiskit code package. 
The gate counts used in Section 2.5 were obtained by summing 
the circuit occurrences of single-qubit rotation U1, U2, and U3 gates and 2-qubit CX gates. These gate 
counts could be further optimized for specific quantum hardware and thus are upper bound estimates for 
fidelity expectations. Parallelization of energy evaluations over multiple processes was performed by 
using the mpi4py and Qiskit code packages. Qiskit was used to construct the appropriate shared quantum 
circuit for each system while an mpi4py wrapper distributed the circuit and one or more needed 
measurement bases equally to the different available processors. Each processor then determines the bit 
string counts that result from each circuit measurement while mpi4py aggregates all resulting data in 
order to calculate the final energy evaluation for a given set of parameters. This code used in this work is available on request and  will be released open-source.

\section*{Funding Information}
This work was authored in part by the National Renewable Energy Laboratory (NREL), operated by Alliance 
for Sustainable Energy, LLC, for the U.S. Department of Energy (DOE) under Contract No. DE-AC36-
08GO28308. This work was supported by the Laboratory Directed Research and Development (LDRD) Program at 
NREL. The views expressed in the article do not necessarily represent the views of the DOE or the U.S. 
Government. The U.S. Government retains and the publisher, by accepting the article for publication, 
acknowledges that the U.S. Government retains a nonexclusive, paid-up, irrevocable, worldwide license to 
publish or reproduce the published form of this work, or allow others to do so, for U.S. Government 
purposes. 


\section{Research Resources}
This research was performed using the Eagle supercomputing resources located at the National Renewable 
Energy Laboratory and sponsored by the DOE’s Office of Energy Efficiency and Renewable Energy.

\selectlanguage{english}
\nocite{*}
\bibliography{VQE4TiH}

\renewcommand\thefigure{S\arabic{figure}}    
\section{Supplemental Information}
\setcounter{figure}{0}   
The following figures supplement those in the main text.  Figures \ref{fig:ccsdAH} through \ref{fig:ccsdProj} concern basis set choice.
Figure \ref{fig:fidelitysupp} is a fidelity plot for an additional combination of method (UCCSD) and basis set choice.
Finally, Figure \ref{fig:gaussians} shows the number of Gaussian primitive functions required as a function of basis set.
\begin{figure}
\centering
\begin{tabular}{cc}
\subf{\includegraphics[width=50mm]{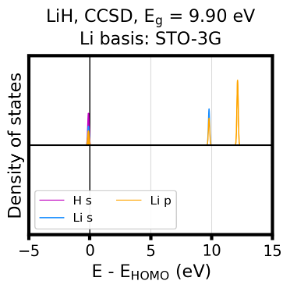}}
     {}
&
\subf{\includegraphics[width=50mm]{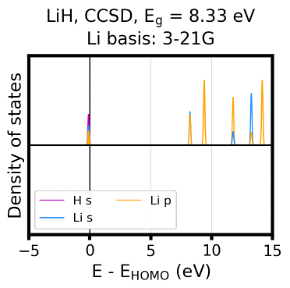}}
     {}
\\
\subf{\includegraphics[width=50mm]{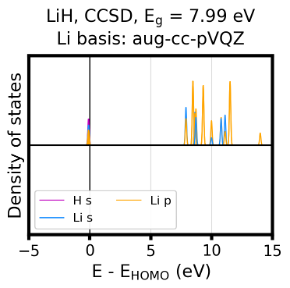}}
     {}
&
\subf{\includegraphics[width=50mm]{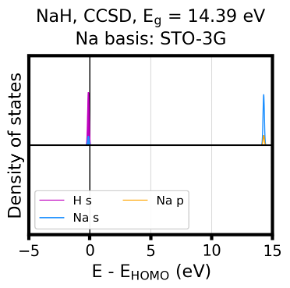}}
     {}
\\
\subf{\includegraphics[width=50mm]{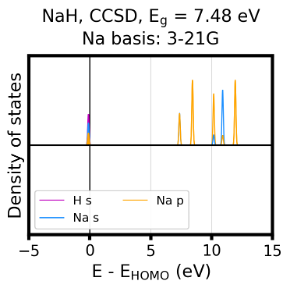}}
     {}
&
\subf{\includegraphics[width=50mm]{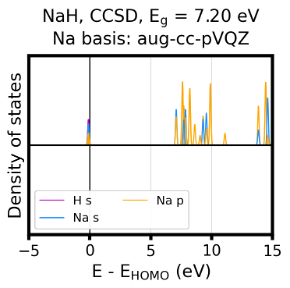}}
     {}
\\
\subf{\includegraphics[width=50mm]{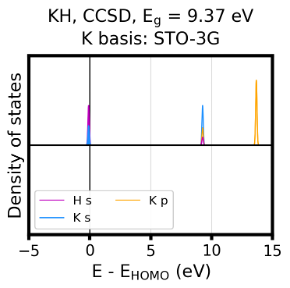}}
     {}
&
\subf{\includegraphics[width=50mm]{ccsdG.png}}
     {}
\\
\end{tabular}
\caption{Classical CCSD projected density of states for LiH, NaH, and KH using different basis sets. The aug-cc-pVQZ PDOS for KH were omitted because no correlation consistent basis sets for K are available within the Gaussian 16 package.}
\label{fig:ccsdAH}
\end{figure}

\begin{figure}
        \centering
            \includegraphics[width=0.5\linewidth]{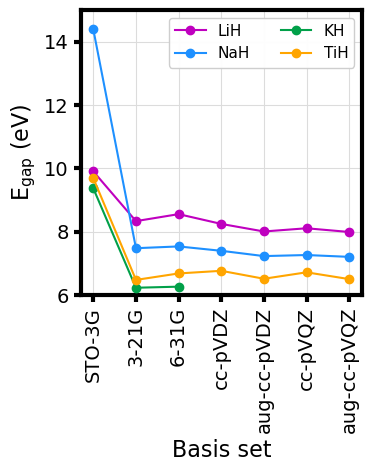}
        \caption{  Classical CCSD HOMO-LUMO gaps of LiH, NaH, KH, and TiH using different basis sets. The correlation consistent gaps for KH were omitted because no correlation consistent basis sets for K are available within the Gaussian 16 package. }
        \label{fig:ccsdHomoLumo}
\end{figure}

\begin{figure}
\centering
\begin{tabular}{cc}
\subf{\includegraphics[width=60mm]{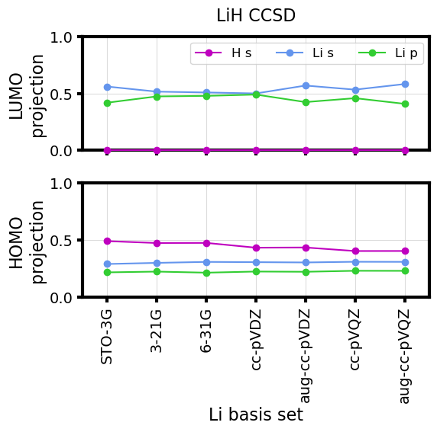}}
     {}
&
\subf{\includegraphics[width=60mm]{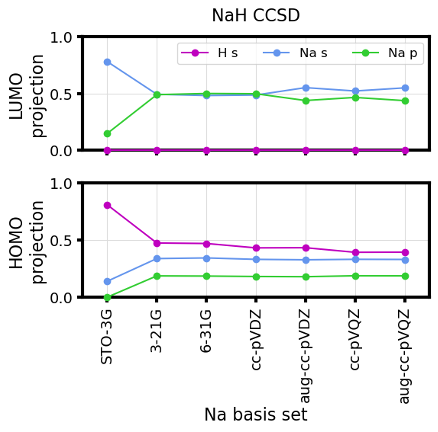}}
     {}
\\
\subf{\includegraphics[width=60mm]{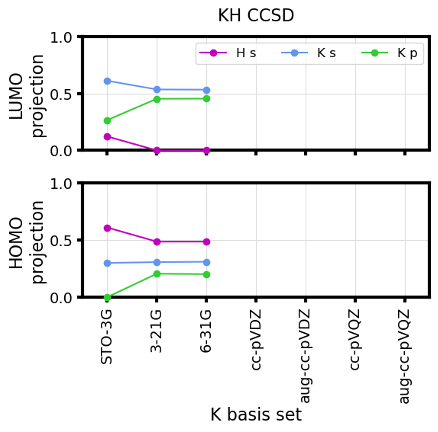}}
     {}
&
\subf{\includegraphics[width=60mm]{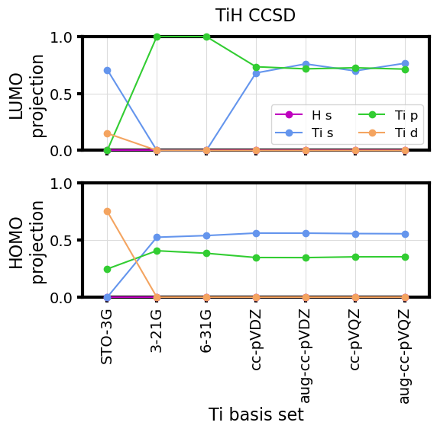}}
     {}
\\
\end{tabular}
\caption{Classical CCSD projected character of the frontier orbitals for LiH, NaH, KH, and TiH using different basis sets. The correlation consistent gaps for KH were omitted because no correlation consistent basis sets for K are available within the Gaussian 16 package.}
\label{fig:ccsdProj}
\end{figure}

\begin{figure}
        \centering
            \includegraphics[scale=0.7]{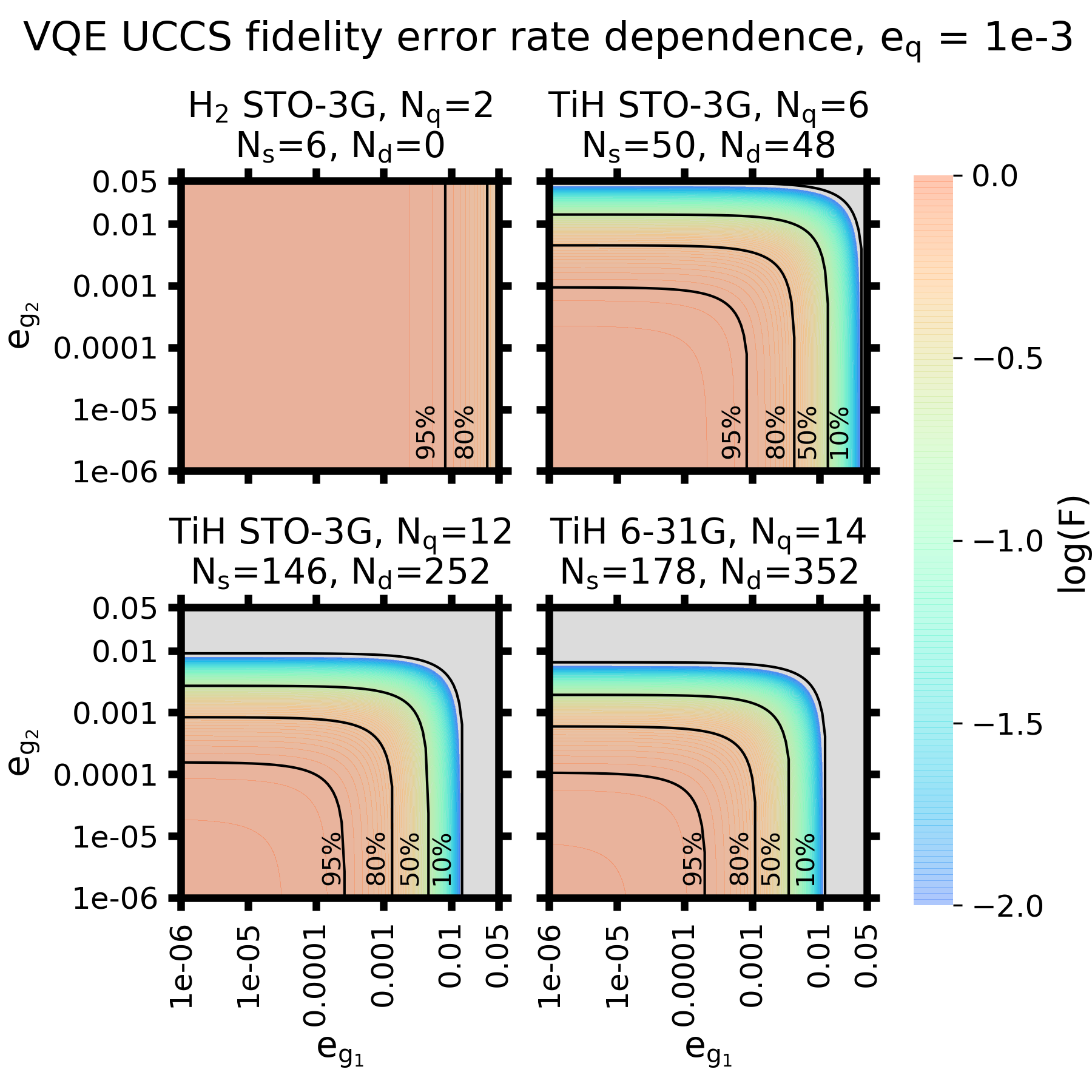}
        \caption{Fidelity estimates for the UCCSD ansatz for the hydride diatomic molecules studied in this work with different numbers of qubits, single qubit gates, and two-qubit gates in the quantum circuit. A fidelity of 1.0 corresponds to no fidelity loss due to the considered errors.}
        \label{fig:fidelitysupp}
\end{figure}

\begin{figure}
        \centering
            \includegraphics[width=0.5\linewidth]{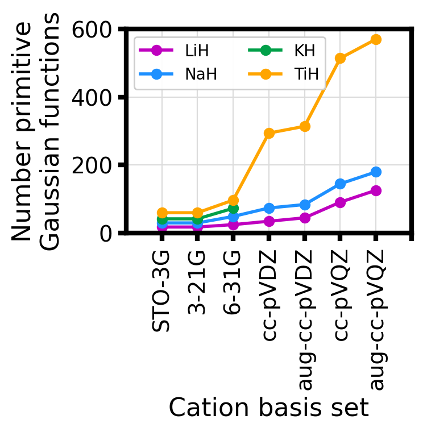}
        \caption{ Number of primitive Gaussian functions used for the classical calculations in Section \ref{sec:hydride}    }
        \label{fig:gaussians}
\end{figure}

\end{document}